\documentclass[10pt]{article}
\usepackage{multicol}
\usepackage{graphicx}
\usepackage{amsmath}
\usepackage[a4paper]{geometry}
\usepackage{rotating}

\begin{document}

\begin{center}
{\Large\bf Chemical potentials of light flavor quarks from yield
ratios of negative to positive particles in Au+Au collisions at
RHIC}

\vskip0.75cm

Ya-Qin Gao$^{1,}${\footnote{E-mail: gyq610@163.com;
gaoyaqin@tyust.edu.cn}}, Hai-Ling Lao$^2$, Fu-Hu
Liu$^{2,}${\footnote{E-mail: fuhuliu@163.com; fuhuliu@sxu.edu.cn}}

\vskip0.2cm

{\small\it $^{1}$Department of Physics, Taiyuan University of
Science and Technology, Taiyuan, Shanxi 030024, China

$^2$Institute of Theoretical Physics \& State Key Laboratory of
Quantum Optics and Quantum Optics Devices,

Shanxi University, Taiyuan, Shanxi 030006, China}

\end{center}

\vskip0.5cm

{\bf Abstract:} The transverse momentum spectra of $\pi^{-}$,
$\pi^{+}$, $K^{-}$, $K^{+}$, $\bar{p}$, and $p$ produced in Au+Au
collisions at center-of-mass energy $\sqrt{s_{NN}}=7.7$, 11.5,
19.6, 27, 39, 62.4, 130, and $200$ GeV are analyzed in the
framework of a multisource thermal model. The experimental data
measured at midrapidity by the STAR Collaboration are fitted by
the (two-component) standard distribution. The effective
temperature of emission source increases obviously with the
increase of the particle mass and the collision energy. At
different collision energies, the chemical potentials of up, down,
and strange quarks are obtained from the antiparticle to particle
yield ratios in given transverse momentum ranges available in
experiments. With the increase of logarithmic collision energy,
the chemical potentials of light flavor quarks decrease
exponentially.
\\



\vskip1.0cm

\begin{multicols}{2}

{\section{Introduction}}

The constructions of the Relativistic Heavy Ion Collider (RHIC)
and the Large Hadron Collider (LHC) have been opening a new epoch
for the studies of nuclear and quark matters. One of the major
goals of the RHIC and LHC studies is to obtain information on the
quantum chromodynamics (QCD) phase diagram~\cite{1}. The phase
diagram includes at least a fundamental phase transition between
the hadron gas and the quark-gluon plasma (QGP) or quark matter,
and is usually plotted as chemical freeze-out temperature
($T_{ch}$) versus baryon chemical potential ($\mu_{baryon}$).
Nowadays, the detailed characteristics of the phase diagram are
not known yet. The experimental and theoretical nuclear physicists
have been focusing their attentions on the searching for the
critical end point and phase boundary. Lattice QCD calculations
show that a system is produced at small $\mu_{baryon}$ or high
energies through a crossover at the quark-hadron phase
transition~\cite{2,3,4}. Based on the lattice QCD~\cite{5} and
several QCD-based models calculations~\cite{6,7,8,9}, as well as
mathematical extensions of lattice techniques~\cite{10,11,12,13},
researchers suggest that the transition at larger $\mu_{baryon}$
is the first order and the QCD critical end point is existent.

Pinpointing the phase boundary and the critical end point is the
central issue to understand the properties of interacting matter
under extreme conditions and to map the QCD phase diagram. The
matter produced in high-energy heavy-ion collisions provides the
opportunity to search for the phase boundary and the critical end
point~\cite{6,14}. To this end, the STAR Collaboration at the RHIC
has undertaken the first phase of the beam energy scan (BES)
program~\cite{15,16,17}, and starting the second phase from 2018
to 2019~\cite{18}. The program is to vary the collision energy
which enables a search for non-monotonic excitation functions over
a broad domain of the phase diagram. Before looking for an
evidence for the existence of a critical end point and the phase
boundary, it is important to know the ($T_{ch},\mu_{baryon}$)
region of phase diagram one can access. The produced particles
spectra and yield ratios allow us only to infer the values of
$T_{ch}$ and $\mu_{baryon}$~\cite{19}. Furthermore, the bulk
properties such as rapidity density $dN/dy$, mean transverse
momentum $\langle p_{T}\rangle$, particle ratios, and freeze-out
properties may provide an insight into the particle production
mechanisms at BES energies. Therefore, it is very important to
study these bulk properties systematically, which may reveal the
evolution and the changes of the system created in high-energy
heavy-ion collisions.

As one of the most important measured quantities, the transverse
momentum ($p_T$) spectrum includes abundant information which are
related to the excitation degree of the collision system. The
spectra of identified particles can also provide useful
information about temperature, particle ratio, and chemical
potential by using thermal and statistical
investigations~\cite{20}. For any system, one can determine the
direction and limitation of mass transfer by comparing the
chemical potentials of particles, that is to say that the chemical
potential is a sign to mark the direction of spontaneous chemical
reaction. The chemical potential can also be a criterion for
determining whether thermodynamic equilibrium does exist in the
interacting region in high-energy collisions~\cite{1}. Generally,
a low absolute value of chemical potential corresponds to a high
degree of thermodynamic equilibrium. Therefore, the chemical
potential is also one of the major solutions for investigating the
QGP. One can see that the chemical potentials of quarks are an
important subject at high energy. Therefore, we are very
interested in measurements the chemical potentials of quarks.

In this paper, we extract the chemical potentials of light flavor
quarks from the yield ratios of negatively to positively charged
particles. By using the (two-component) standard distribution, the
$p_T$ spectra of $\pi^{-}$, $\pi ^{+}$, $K^{-}$, $K^{+}$,
$\bar{p}$, and $p$ produced in Au+Au collisions at center-of-mass
energy (per nucleon pair) $\sqrt{s_{NN}}=7.7$, 11.5, 19.6, 27, 39,
62.4, 130, and 200 GeV measured by the STAR Collaboration in
midrapidity interval ($|y|<0.1$)~\cite{19,21} are described. The
considered energies stretch across a wide energy range which
covers the main range of the RHIC at its BES.
\\

{\section{The model and method}}

To extract the chemical potentials of quarks, we need to know the
yield ratios of negatively to positively charged particles.
Although we can have the values of yield ratios directly in
experiments, they are not complete and comprehensive in some
cases. Usually, the $p_T$ spectra of charged particles are given
in many experiments and we can get the yield ratios by fitting the
available data. Then, the values of chemical potentials for the
up, down, and strange quarks can be obtained from the yield ratios
$\pi^{-}/\pi^{+}$, $K^{-}/K^{+}$, and $\bar{p}/p$ which are
synthetically considered in special ways.

In this paper, the $p_T$ spectra are analyzed in the framework of
a multisource thermal model~\cite{22}, which assumes that various
sources are involved in high-energy collisions. These sources are
divided into few groups by different interaction mechanisms,
geometrical relations, or event samples. Each group of sources
forms a relatively large emission source which stays in a local
thermal equilibrium state at the chemical or kinetic freeze-out.
Each emission source is considered to emit particles in its rest
frame and treated as a thermodynamic system of relativistic and
quantum ideal gas. This means that each emission source can be
described by the thermal and statistical model or other similar
models and distributions. The final-state distribution is
attributed to all sources in the whole system, which results in a
multi-characteristic emission process~\cite{22} if we use the
standard distribution~\cite{23,24,25,26}. This also means that
$p_T$ spectrum can be described by a multi-component standard
distribution in which each component describes a given emission
source.

We now structure the multi-component standard distribution. It is
assumed that there are $l$ components to be considered. For the
$i$-th component, the standard Boltzmann, Fermi-Dirac and
Bose-Einstein distributions~\cite{23,24,25,26} can be uniformly
expressed as
\begin{align}
f_{i}\left( p_{T}\right) = & \frac{1}{N} \frac{dN}{dp_{T}} =
C_{i}p_{T}\sqrt{p_{T}^{2}+m_{0}^{2}} \int_{y_{\min}}^{y_{\max}}
\cosh y \nonumber\\
& \times \left[ \exp\left( \frac{\sqrt{p_{T}^{2}+m_{0}^{2}}\cosh
y-\mu}{T_{i}}\right) +S\right] ^{-1}dy,
\end{align}
where $C_{i}$ is the normalization constant which results in
$\int_{0}^{\infty }f_{i}(p_{T})dp_T=1$; $N$, $m_{0}$, $\mu$, and
$T_{i}$ denote the particle number, the rest mass of the
considered particle, the chemical potential of the considered
particle, and the effective temperature for the $i$-th component,
respectively; $y_{\min}$ is the minimum rapidity and $y_{\max}$ is
the maximum rapidity; the values of $S$ are 0, $+1$, and $-1$,
which denote the Boltzmann, Fermi-Dirac, and Bose-Einstein
distributions, respectively. We neglect the existence of $\mu$ in
Eq. (1) due to the fact that it has mainly effect on the
normalization which can be redone, but not the trend of curve.

In the final state, $p_T$ spectrum is resulting from $l$
components, that is
\begin{equation}
f\left( p_{T}\right) =\frac{1}{N}\frac{dN}{dp_{T}}=
{\displaystyle\sum\limits_{i=1}^{l}} w_{i}f_{i}\left(
p_{T}\right),
\end{equation}
where $w_{i}$ ($i=1$, 2, $\cdots$, $l$) is the relative weight
resulting from the $i$-th component. Because of the probability
distribution being acquiescently normalized to 1, the coefficient
obeys the normalization condition of $\sum w_{i}=1$. Considering
the relative contribution of each component, we have the mean
effective temperature to be $T_{eff}= \sum w_{i}T_{i}$, which
reflects the mean excitation degree of different sources
corresponding to different components and can be used to describe
the effective temperature of whole interacting system. It should
be noted that the effective temperature contains the contributions
of transverse flow and thermal motion. It is not the ``real"
temperature of the interacting system.

According to Refs.~\cite{27,28}, the relation between antiproton
to proton yield ratios can be written as
\begin{equation}
\frac{\bar{p}}{p} =\exp\left( -\frac{2\mu_{p}}{T_{ch}}\right)
\approx\exp\left( -\frac{2\mu_{baryon}}{T_{ch}}\right),
\end{equation}
where $\mu_p$ denotes the chemical potential of proton. In the
framework of the statistical thermal model of non-interacting gas
particles with the assumption of standard Maxwell-Boltzmann
statistics, there is an empirical expression for
$T_{ch}$~\cite{29,30,31,32}, one has
\begin{equation}
T_{ch}=T_{\lim}\frac{1}{1+\exp\left[ 2.60-\ln\left( \sqrt{s_{NN}}
\right)/0.45\right]},
\end{equation}
where $\sqrt{s_{NN}}$ is in the units of GeV and the ``limiting"
temperature $T_{\lim}=0.164$ GeV~\cite{29,30}.

In a similar way, the yield ratios of antiparticles to particles
for other hadrons can be written as
\begin{align}
k_{\pi}&\equiv\frac{\pi^{-}}{\pi^{+}} =\exp\left(-\frac{2\mu_{\pi}}{T_{ch}}\right),\nonumber\\
k_{K}&\equiv\frac{K^{-}}{K^{+}} =\exp\left(-\frac{2\mu_{K}}{T_{ch}}\right),\nonumber\\
k_{p}&\equiv\frac{\bar{p}}{p} =\exp\left(-\frac{2\mu_{p}}{T_{ch}}\right),\nonumber\\
k_{D}&\equiv\frac{D^{-}}{D^{+}} =\exp\left(-\frac{2\mu_{D}}{T_{ch}}\right),\nonumber\\
k_{B}&\equiv\frac{B^{-}}{B^{+}}
=\exp\left(-\frac{2\mu_{B}}{T_{ch}}\right),
\end{align}
where $k_{j}$ ($j=\pi$, $K$, $p$, $D$, and $B)$ denote the yield
ratios of negatively to positively charged particles obtained from
the normalization constants of $p_T$ spectra. The symbols
$\mu_{\pi}$, $\mu_{K}$, $\mu_{D}$, and $\mu_{B}$ represent the
chemical potentials of $\pi$, $K$, $D$, and $B$, respectively. In
the above discussion, the symbol of a given particle is used for
its yield for the purpose of simplicity. Furthermore, we have
\begin{align}
\mu_{\pi} & =-\frac{1}{2}T_{ch}\cdot\ln\left( k_{\pi}\right),\nonumber\\
\mu_{K} & =-\frac{1}{2}T_{ch}\cdot\ln\left( k_{K}\right),\nonumber\\
\mu_{p} & =-\frac{1}{2}T_{ch}\cdot\ln\left( k_{p}\right),\nonumber\\
\mu_{D} & =-\frac{1}{2}T_{ch}\cdot\ln\left( k_{D}\right),\nonumber\\
\mu_{B} & =-\frac{1}{2}T_{ch}\cdot\ln\left( k_{B}\right).
\end{align}

Let $\mu_{q}$ denote the chemical potential for quark flavor,
where $q=u$, $d$, $s$, $c$, and $b$ represent the up, down,
strange, charm, and bottom quarks, respectively. In principle, we
can use $k_j$ to give relations among different $\mu_{q}$. The
values of $\mu_{q}$ are then expected from these relations.
According to Refs.~\cite{33,34}, based on the same $T_{ch}$, $k_j$
in terms of $\mu_q$ are
\begin{align}
k_{\pi} & =\exp\left[ -\frac{\left( \mu_{u}-\mu_{d}\right)}
{T_{ch}}\right] \bigg/\exp\left[  \frac{\left(
\mu_{u}-\mu_{d}\right)} {T_{ch}}\right] \nonumber\\ & =\exp\left[
-\frac{2\left(\mu_{u}-\mu_{d}\right)}
{T_{ch}}\right],\nonumber\\
k_{K} & =\exp\left[ -\frac{\left( \mu_{u}-\mu_{s}\right)}
{T_{ch}}\right] \bigg/\exp\left[ \frac{\left(
\mu_{u}-\mu_{s}\right)} {T_{ch}}\right] \nonumber\\ & =\exp\left[
-\frac{2\left( \mu_{u}-\mu_{s}\right)}
{T_{ch}}\right],\nonumber\\
k_{p} & =\exp\left[ -\frac{\left( 2\mu_{u}+\mu_{d}\right)}
{T_{ch}}\right] \bigg/\exp\left[ \frac{\left(
2\mu_{u}+\mu_{d}\right)} {T_{ch}}\right] \nonumber\\ & =\exp\left[
-\frac{2\left(2\mu_{u}+\mu_{d}\right)}
{T_{ch}}\right],\nonumber\\
k_{D} & =\exp\left[ -\frac{\left( \mu_{c}-\mu_{d}\right)}
{T_{ch}}\right] \bigg/\exp\left[ \frac{\left(
\mu_{c}-\mu_{d}\right)} {T_{ch}}\right] \nonumber\\ & =\exp\left[
-\frac{2\left( \mu_{c}-\mu_{d}\right)}
{T_{ch}}\right],\nonumber\\
k_{B} & =\exp\left[ -\frac{\left( \mu_{u}-\mu_{b}\right)}
{T_{ch}}\right] \bigg/\exp\left[ \frac{\left(
\mu_{u}-\mu_{b}\right)} {T_{ch}}\right] \nonumber\\ & =\exp\left[
-\frac{2\left( \mu_{u}-\mu_{b}\right)} {T_{ch}}\right].
\end{align}
Thus, we have
\begin{align}
\mu_{u} & =-\frac{1}{6}T_{ch}\cdot \ln\left( k_{\pi}\cdot
k_{p}\right),\nonumber\\
\mu_{d} & =-\frac{1}{6}T_{ch}\cdot \ln\left( k_{\pi}^{-2}\cdot
k_{p}\right),\nonumber\\
\mu_{s} & =-\frac{1}{6}T_{ch}\cdot \ln\left( k_{\pi}\cdot
k_{K}^{-3}\cdot k_{p}\right),\nonumber\\
\mu_{c} & =-\frac{1}{6}T_{ch}\cdot \ln\left( k_{\pi}^{-2}\cdot
k_{p}\cdot k_{D}^{3}\right),\nonumber\\
\mu_{b} & =-\frac{1}{6}T_{ch}\cdot \ln\left( k_{\pi}\cdot
k_{p}\cdot k_{B}^{-3}\right).
\end{align}

As can be seen from Eq. (8) that $\mu_q$ are obtained from $k_j$.
In addition to the yield ratios $\pi^{-}/\pi^{+}$, $K^{-}/K^{+}$
and $\bar{p}/p$, other combinations can also give $\mu_{q}$ if the
spectra in the numerator and denominator are under the same
experimental conditions.
\\

{\section{Results and discussion}}

\begin{figure*}
\hskip-1.cm
\includegraphics[width=20.cm]{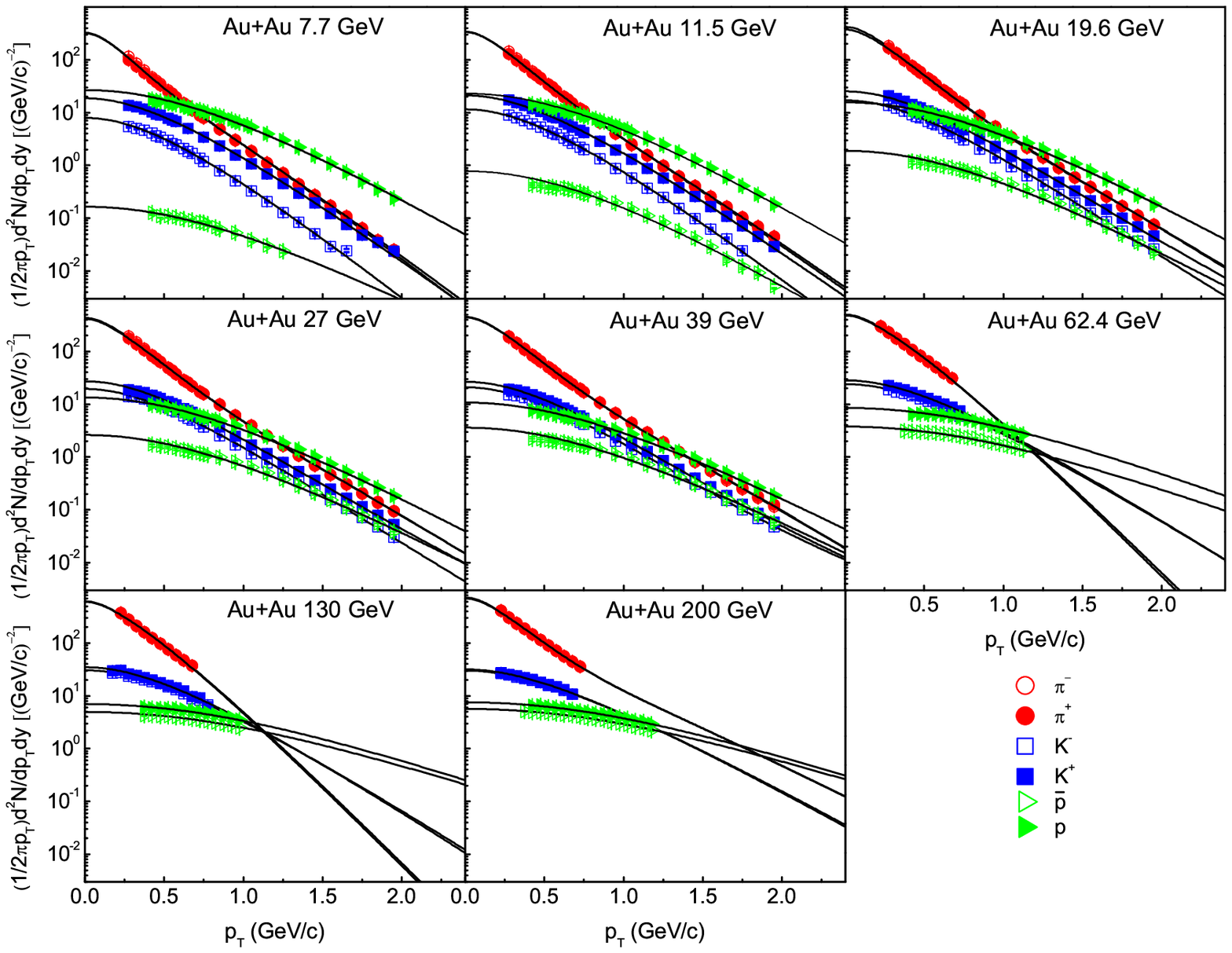}
Fig. 1: Midrapidity ($|y|<0.1$) double-differential $p_T$ spectra
for $\pi^{-}$, $\pi^{+}$, $K^{-}$, $K^{+}$, $\bar{p}$, and $p$ in
central Au+Au collisions at $\sqrt{s_{NN}}=7.7$, 11.5, 19.6, 27,
39, 62.4, 130, and 200 GeV, where the centrality interval at 130
GeV is 0--6\% and at other energies is 0--5\%. The different
symbols represent the measurements done by the STAR
experiment~\cite{19,21} and the curves represent the results
fitted by the (two-component) standard distribution. The values of
parameters can be found in Table 1.
\end{figure*}

The energy dependent double-differential $p_T$ spectra of $\pi^-$,
$\pi^+$, $K^-$, $K^+$, $\bar{p}$, and $p$ produced in central
Au+Au collisions at $\sqrt{s_{NN}}=7.7$, 11.5, 19.6, 27, 39, 62.4,
130, and 200 GeV at the midrapidity $|y|<0.1$ are presented in
Fig. 1, where the centrality interval at 130 GeV is 0--6\% and at
other energies is 0--5\%. The different symbols represent the data
measured by the STAR Collaboration~\cite{19,21}, and the curves
are the results fitted here by the (two-component) standard
distribution. Generally, the standard distribution is firstly used
in the fit process. If it does not fit the data, the two-component
standard distribution is used. It is because of the quality of the
measurements that (two-component) standard distribution is used.
In the case of using the two-component standard distribution, the
first component results in narrow $p_T$ region and the second
component results in wide $p_T$ regions. That is, in low $p_T$
region both components contribute to the spectra, and in high
$p_T$ region only the second component contributes to the spectra.
In the calculation, the values of the free parameters ($T_1$,
$w_1$, and $T_2$), the normalization constant ($N_0$), and
$\chi^2$ obtained by fitting the data are listed in Table 1
including the degrees of freedom (dof). One can see that the data
are well fitted by the (two-component) standard distribution. From
the parameter values, one can see that the effective temperature
increases with the increase of the particle mass and the collision
energy for emissions of the six types of particles.

Based on the above successful fits of the $p_T$ spectra of
antiparticles and particles, we can use Eq. (8) and the $p_T$
spectra in Fig. 1 to study the dependence of $\mu_q$ on
$\sqrt{s_{NN}}$. This is done by integrating the yield over the
given $p_T$ ranges available in experiments at different energies.
Figure 2 shows the correlations between $\mu_q$ and
$\sqrt{s_{NN}}$, where $\mu$ is used on the vertical axis to
replace $\mu_q$ which are marked in the panel for different styles
of symbols. With the increase of logarithmic $\sqrt{s_{NN}}$, an
exponential decrease of $\mu_q$ is observed. Corresponding to the
solid, dashed and dotted curves which fit to the dependences of
$\mu_u$, $\mu_d$, and $\mu_s$ on $\sqrt{s_{NN}}$, respectively, we
have
\begin{align}
\mu_{u}&=(820.1\pm0.1) \big(\sqrt{s_{NN}}\hspace{.5mm}{\rm [GeV]}\big)^{-(0.914\pm0.025)}\hspace{1.2mm}{\rm MeV},\nonumber\\
\mu_{d}&=(681.1\pm0.1) \big(\sqrt{s_{NN}}\hspace{.5mm}{\rm [GeV]}\big)^{-(0.834\pm0.031)}\hspace{1.2mm}{\rm MeV},\nonumber\\
\mu_{s}&=(420.7\pm0.2) \big(\sqrt{s_{NN}}\hspace{.5mm}{\rm
[GeV]}\big)^{-(1.004\pm0.063)}\hspace{1.2mm}{\rm MeV}
\end{align}
with the $\chi^{2}/{\rm dof}=1.67/6$, 2.73/6, and 1.04/6,
respectively.

The similarity in up and down quark masses renders the similarity
in their chemical potentials. The difference between the chemical
potentials of up (or down) and strange quarks is caused by the
difference between their masses. At the lowest BES energy the
difference between the chemical potentials are dozens of MeV,
while at the highest RHIC energy these quantities are around a few
MeV. The decrease in $\mu_q$ is obvious, which indicates the
change of mean free path of produced quarks in the middle state.
If the produced quarks at the lowest BES energy have a small mean
free path which looks as if a liquid-like middle state is formed,
the produced quarks at the highest RHIC energy should have a large
mean free path which looks as if a gas-like middle state is
formed. The main difference at different energies is different
mean free paths of the produced quarks. To search for the critical
energy at which the change from a liquid-like middle state to a
gas-like middle state had happen is beyond the focus of the
present work.

\end{multicols}
\begin{sidewaystable}
{Table 1: Values of $T_{1}$, $w_{1}$, $T_{2}$, $N_0$, $\chi^{2}$,
and dof corresponding to the curves in Fig. 1. The values for
positively and negatively charged particles are given using the
slash (/), where the values for positive particles are shown
before the slash and the values for negative particles after the
slash. The values of $N_0$ are obtained due to the comparisons
between the experimental $(1/2\pi p_T)d^2N/dp_Tdy$ and the
calculated $(1/2\pi p_T)N_0f(p_T)/dy$, where $dy=0.2$ and $f(p_T)$
is presented in Eq. (2).}
\begin{center}
{\small
\begin{tabular}{cccccccc}
\hline\hline $\sqrt{s_{NN}}$ (GeV) & Particles & $T_{1}$ (MeV) &
$w_{1}$ & $T_{2}$ (MeV) & $N_0$ & $\chi ^{2}$ & dof \\
\hline
 7.7  & $\pi^{\pm}$   & $ 87\pm 7/ 89\pm10$ & $0.65\pm0.04/0.63\pm0.07$ & $181\pm 2/177\pm 4$ & $81.0\pm4.0/81.0\pm9.0$     & 0.38/0.45  & 22/22 \\
      & $K^{\pm}$     & $185\pm 6/170\pm 8$ & 1.00/1.00                 & $-$                 & $9.5\pm0.8/3.8\pm0.5$       & 0.75/0.98  & 21/21 \\
      & $p$/$\bar{p}$ & $224\pm 7/257\pm20$ & 1.00/1.00                 & $-$                 & $19.0\pm0.2/0.13\pm0.01$    & 1.15/0.52  & 27/13 \\ \hline
 11.5 & $\pi^{\pm}$   & $102\pm10/105\pm10$ & $0.70\pm0.04/0.69\pm0.03$ & $193\pm 4/190\pm 4$ & $92.0\pm8.0/93.0\pm8.0$     & 0.25/0.48  & 22/22 \\
      & $K^{\pm}$     & $190\pm 5/178\pm 4$ & 1.00/1.00                 & $-$                 & $11.0\pm0.8/5.7\pm0.5$      & 0.49/0.97  & 23/21 \\
      & $p$/$\bar{p}$ & $217\pm10/216\pm20$ & 1.00/1.00                 & $-$                 & $16.0\pm3.0/0.54\pm0.10$    & 1.70/3.07  & 26/21 \\ \hline
 19.6 & $\pi^{\pm}$   & $118\pm10/116\pm10$ & $0.79\pm0.02/0.83\pm0.02$ & $215\pm 8/219\pm 7$ & $109.0\pm10.0/119.0\pm10.0$ & 0.19/0.40  & 22/22 \\
      & $K^{\pm}$     & $181\pm 7/181\pm 5$ & $0.88\pm0.04/0.89\pm0.05$ & $260\pm20/239\pm20$ & $13.0\pm1.4/8.7\pm0.8$      & 0.55/1.61  & 22/22 \\
      & $p$/$\bar{p}$ & $234\pm20/237\pm10$ & 1.00/1.00                 & $-$                 & $11.5\pm2.0/1.4\pm0.1$      & 0.69/19.81 & 27/20 \\ \hline
 27   & $\pi^{\pm}$   & $117\pm 8/118\pm 7$ & $0.79\pm0.03/0.81\pm0.03$ & $219\pm 7/221\pm 7$ & $120.0\pm10.0/125.0\pm10.0$ & 0.19/0.27  & 22/22 \\
      & $K^{\pm}$     & $178\pm 7/180\pm 6$ & $0.85\pm0.03/0.84\pm0.06$ & $262\pm20/236\pm10$ & $14.0\pm1.0/10.0\pm1.0$     & 0.27/0.92  & 22/21 \\
      & $p$/$\bar{p}$ & $239\pm 7/247\pm10$ & 1.00/1.00                 & $-$                 & $10.0\pm2.0/2.0\pm0.2$      & 0.84/1.28  & 21/20 \\ \hline
 39   & $\pi^{\pm}$   & $114\pm10/116\pm10$ & $0.78\pm0.02/0.78\pm0.02$ & $222\pm 7/222\pm 6$ & $129.0\pm12.0/128.0\pm10.0$ & 0.39/0.20  & 22/22 \\
      & $K^{\pm}$     & $189\pm 7/189\pm 5$ & $0.95\pm0.01/0.96\pm0.02$ & $332\pm18/347\pm23$ & $14.0\pm1.5/11.0\pm1.0$     & 0.20/0.36  & 22/22 \\
      & $p$/$\bar{p}$ & $250\pm20/252\pm20$ & 1.00/1.00                 & $-$                 & $8.3\pm1.0/2.8\pm0.3$       & 0.81/2.42  & 20/21 \\ \hline
 62.4 & $\pi^{\pm}$   & $139\pm 6/137\pm 8$ & 1.00/1.00                 & $-$                 & $146.0\pm10.0/150.0\pm10.0$ & 1.14/1.07  & 8/8   \\
      & $K^{\pm}$     & $210\pm25/214\pm25$ & 1.00/1.00                 & $-$                 & $15.8\pm1.0/13.6\pm0.8$     & 0.02/0.37  & 8/8   \\
      & $p$/$\bar{p}$ & $335\pm20/346\pm40$ & 1.00/1.00                 & $-$                 & $8.3\pm0.4/3.8\pm0.3$       & 1.97/1.60  & 13/14 \\ \hline
 130  & $\pi^{\pm}$   & $136\pm 8/137\pm 7$ & 1.00/1.00                 & $-$                 & $181.0\pm13.0/185.0\pm11.0$ & 1.98/3.68  & 8/8   \\
      & $K^{\pm}$     & $204\pm12/210\pm13$ & 1.00/1.00                 & $-$                 & $18.5\pm1.4/17.3\pm1.3$     & 0.39/0.20  & 11/11 \\
      & $p$/$\bar{p}$ & $373\pm15/385\pm22$ & 1.00/1.00                 & $-$                 & $7.5\pm0.7/5.5\pm0.3$       & 1.73/0.47  & 11/11 \\ \hline
 200  & $\pi^{\pm}$   & $116\pm 8/115\pm 5$ & $0.76\pm0.03/0.76\pm0.03$ & $263\pm23/262\pm25$ & $209.0\pm14.0/215.0\pm12.0$ & 0.21/0.17  & 7/7   \\
      & $K^{\pm}$     & $235\pm30/239\pm30$ & 1.00/1.00                 & $-$                 & $19.3\pm1.1/18.5\pm1.0$     & 0.06/0.06  & 8/8   \\
      & $p$/$\bar{p}$ & $382\pm40/393\pm35$ & 1.00/1.00                 & $-$                 &$8.3\pm0.9/6.4\pm0.6$        & 0.10/0.25  & 14/15 \\
\hline
\end{tabular}}
\end{center}
\end{sidewaystable}
\begin{multicols}{2}

\begin{figure*}
\vskip-1.0cm
\begin{center}
\includegraphics[width=14.0cm]{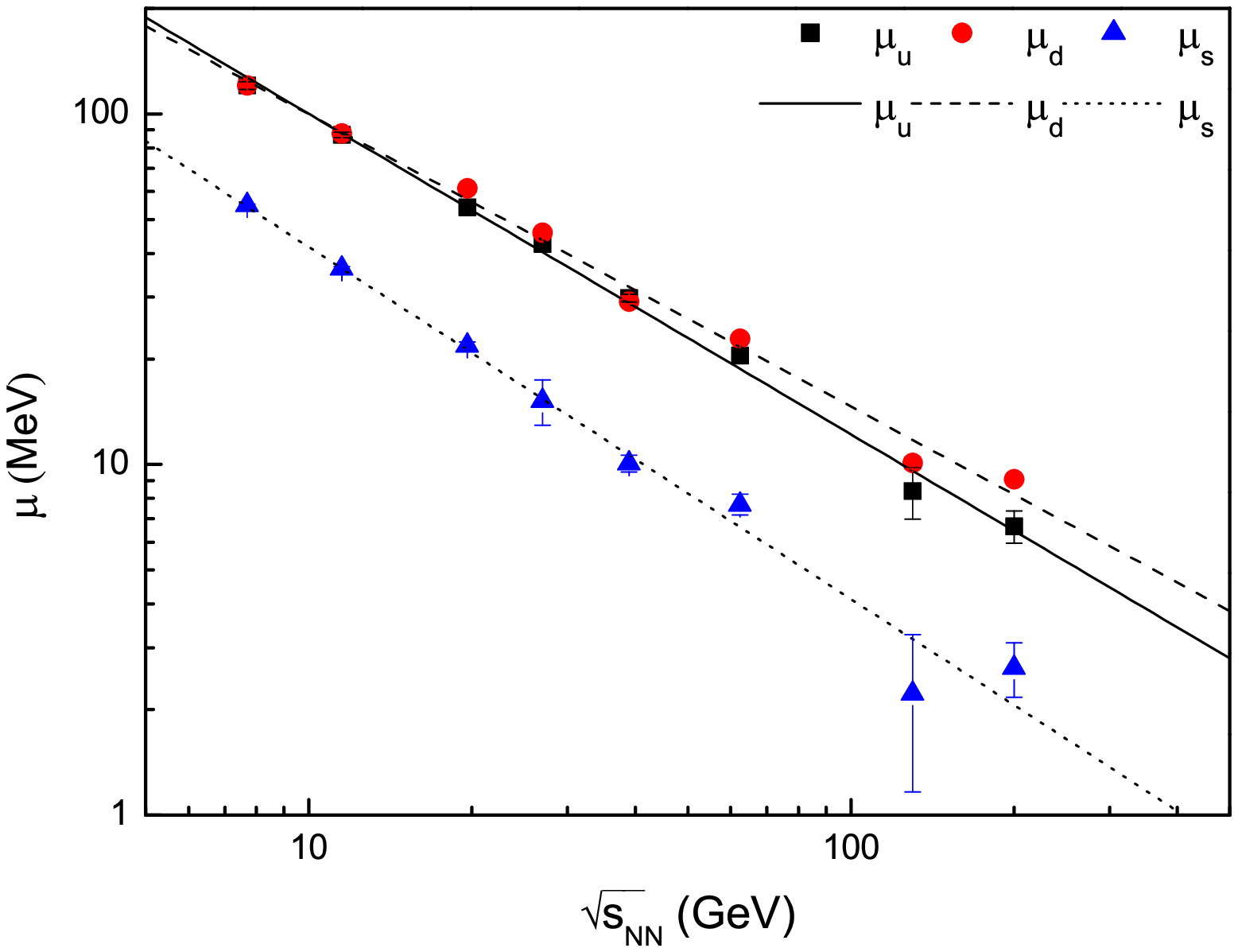}
\end{center}
Fig. 2: Correlations between $\mu_q$ and $\sqrt{s_{NN}}$ for
central Au+Au collisions at RHIC. The symbols represent $\mu_q$
obtained from the ratios by integrating the yield over the given
$p_T$ ranges available in experiments in Fig. 1. The solid, dashed
and dotted curves are fitted results corresponding to $\mu_u$,
$\mu_d$ and $\mu_s$, respectively.
\end{figure*}

From Eq. (9) we can obtain a linear relation between $\ln\mu_q$
and $\ln\sqrt{s_{NN}}$,
\begin{align}
\ln\mu_q = a-|b|\cdot\ln\sqrt{s_{NN}},
\end{align}
where the intercept $a$ and slope $-|b|$ can be obtained from the
parameters in Eq. (9). In particular, $-|b|$ is close to $-1$. The
large negative slope shows an obvious anticorrelation between
$\ln\mu_q$ and $\ln\sqrt{s_{NN}}$. It is expected that $\ln\mu_q$
will be smaller at higher energy or larger at lower energy. In
particular, at the LHC energies, $\ln\mu_q$ will be negative since
$\mu_q$ will be less than 1 MeV. The limiting value of $\mu_q$ is
close to 0 at the LHC [35], which results in an obvious negative
$\ln\mu_q$.

The main conclusion observed from Fig. 2 is that $\mu_q$ is high
(from dozens of MeV to $\sim$100 MeV) at the BES and close to 0 at
the LHC~\cite{35}. This is consistent with the trend of
$\mu_{baryon}$ ($\sim$100--300 MeV at the BES and $\sim$1 MeV at
the LHC) obtained from other works~\cite{1,29,30,31,32,36}. This
is natural due to the fact that baryon is consisted of valence
quarks. If we regard $\mu_{baryon}=\Sigma\mu_q$, where $\Sigma$
denotes the sum over all valence quarks in baryon, the present
work is consistent with the models which study
$\mu_{baryon}$~\cite{1,29,30,31,32,36}.

We would like to point out that although we have used the
(two-component) standard distribution in the fits of $p_T$ spectra
and $T_1$ ($T_2$) has been used, the values of $\mu_q$ obtained by
us are independent of models and parameters. In fact, $\mu_q$ is
only related to $k_j$ if $T_{ch}$ is known. We can use directly
the yield ratios of data to obtain $\mu_q$. The reason why we use
the function form instead of data is to extend $p_T$ spectrum in
intermediate region to low and high regions where the data are not
available. In our opinion, the function fitted the data in
intermediate $p_T$ region can predict approximately the trends in
low and high $p_T$ regions.
\\

{\section{Conclusions}}

In summary, we found a good fit of the transverse momentum spectra
of charged particles produced in central Au+Au collisions at the
RHIC at its BES energies. It is shown that the (two-component)
standard distribution successfully fitted the data measured at
midrapidity by the STAR Collaboration, though other distributions
are also acceptable. The effective temperature parameter increases
with the increase of the particle mass and the collision energy.

At BES energies, the chemical potentials of light flavor quarks
were obtained from the yield ratios of negatively to positively
charged particles in given transverse momentum ranges available in
experiments. At low energy, the chemical potentials of up and down
quarks are consistent but differ from that of strange quark. At
high energy, the three chemical potentials seem to deviate from
each other, and they finally approach zero at very high energy.

From the lowest BES energy to the highest RHIC energy, with the
increase of logarithmic collision energy, an exponential decrease
of the chemical potentials of light flavor quarks is observed. The
similarity in up and down quark masses renders the similarity in
their chemical potentials. The difference between the chemical
potentials of up (or down) and strange quarks is caused by their
different masses. The difference between the chemical potentials
changes from dozens of MeV to a few MeV. The decrease in chemical
potential indicates that the mean free path of produced quarks
changes from a small value to a large one.
\\

{\bf Data Availability}

All data are quoted from the mentioned references. As a
phenomenological work, this paper does not report new data.
\\

{\bf Conflicts of Interest}

The authors declare that there are no conflicts of interest
regarding the publication of this paper.
\\

{\bf Acknowledgments}

Communications from Edward K. Sarkisyan-Grinbaum are highly
acknowledged. This work was supported by the National Natural
Science Foundation of China under Grant Nos. 11747063, 11575103,
and 11747319, the Doctoral Scientific Research Foundation of
Taiyuan University of Science and Technology under Grant No.
20152043, the Shanxi Provincial Natural Science Foundation under
Grant No. 201701D121005, and the Fund for Shanxi ``1331 Project"
Key Subjects Construction.
\\

\end{multicols}
\end{document}